\documentclass[conference]{IEEEtran}
\usepackage{cite}
\usepackage{amsmath,amssymb,amsfonts}
\usepackage{algorithmic}
\usepackage{graphicx}
\usepackage{textcomp}
\usepackage{xcolor}
\usepackage{url}
\usepackage{mathtools}
\usepackage[inline]{enumitem}
\usepackage{float}
\usepackage{caption}
\usepackage{subcaption}

\def\BibTeX{{\rm B\kern-.05em{\sc i\kern-.025em b}\kern-.08em
    T\kern-.1667em\lower.7ex\hbox{E}\kern-.125emX}}
\begin{document}

\title{Learning-Based Practical Light Field Image Compression Using A Disparity-Aware Model
}

\author{\IEEEauthorblockN{Mohana Singh}
\IEEEauthorblockA{\textit{School of Computing and Electrical Engineering} \\
\textit{Indian Institute of Technology Mandi}\\
\textit{Himachal Pradesh, India}\\
s18002@students.iitmandi.ac.in}
\and
\IEEEauthorblockN{Renu M. Rameshan}
\IEEEauthorblockA{\textit{School of Computing and Electrical Engineering} \\
\textit{Indian Institute of Technology Mandi}\\
\textit{Himachal Pradesh, India}\\
renumr@iitmandi.ac.in}}

\maketitle

\begin{abstract}
Light field technology has increasingly attracted the attention of the research community with its many possible applications. The lenslet array in commercial plenoptic cameras helps capture both the spatial and angular information of light rays in a single exposure. While the resulting high dimensionality of light field data enables its superior capabilities, it also impedes its extensive adoption. Hence, there is a compelling need for efficient compression of light field images. Existing solutions are commonly composed of several separate modules, some of which may not have been designed for the specific structure and quality of light field data. This increases the complexity of the codec and results in impractical decoding runtimes. We propose a new learning-based, disparity-aided model for compression of 4D light field images capable of parallel decoding. The model is end-to-end trainable, eliminating the need for hand-tuning separate modules and allowing joint learning of rate and distortion. The disparity-aided approach ensures the structural integrity of the reconstructed light fields. Comparisons with the state of the art show encouraging performance in terms of PSNR and MS-SSIM metrics. Also, there is a notable gain in the encoding and decoding runtimes. Source code is available at https://moha23.github.io/LF-DAAE.

\end{abstract}


\section{Introduction}
In the rapidly expanding realm of visual communication, 4D light field photography has come up as an attractive imaging technology with a multitude of applications, including novel view synthesis, post-capture digital refocusing \cite{ng2005light}, scene reconstruction and enhanced VR experiences. The enabling factor behind its numerous applications is the additional information captured in such 4D light field images as compared to conventional 2D digital images. However, the sheer volume of light field data causes hindrance to its extensive adoption. Hence, the compression of light field images remains one of the crucial tasks in extending its practicality.

\begin{figure*}[htb]
\begin{center}
\includegraphics[width=18cm]{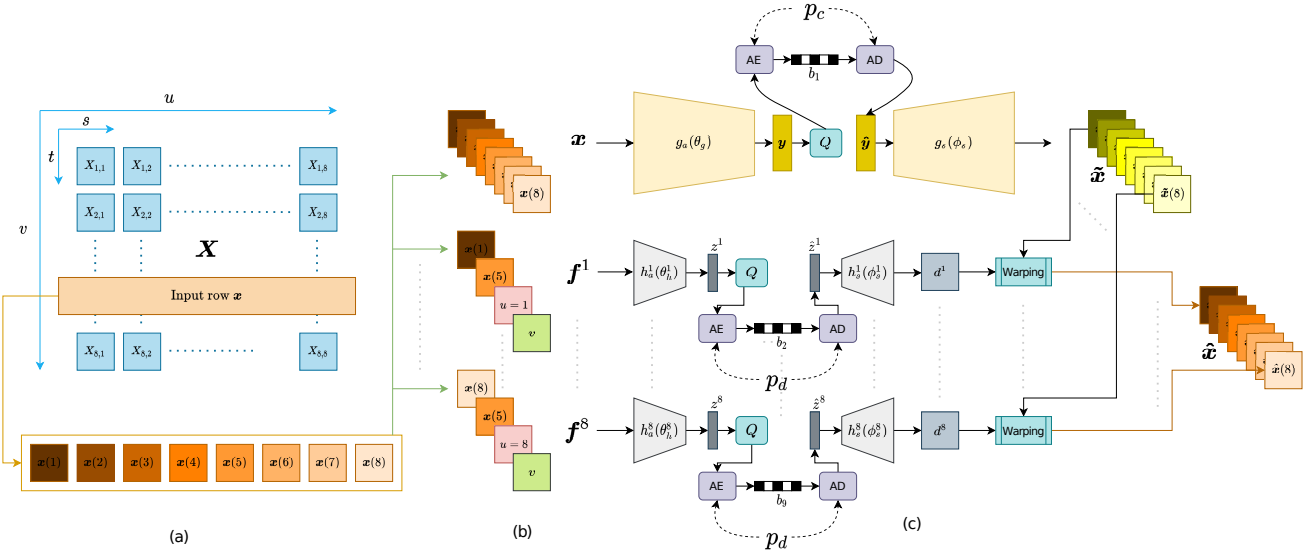}

\caption{(a) Illustration of 4D light field data $\boldsymbol{X}$, where each $X_{i,j}$ is an SAI. (b) Input data is structured by taking one row of SAIs at a time, and creating feature tensors $\boldsymbol{f_{}^{i}}$ using the $(u,v)$ position indices of SAI $\boldsymbol{x}(i)$ (c)  The proposed architecture where each trapezoidal component consists of 3DCNNs and GDN/IGDN nonlinearities. $Q$ is the quantization operation, AE and AD are the arithmetic encoder and decoders, respectively. The final bitstream is composed of each of the $b_{i}\text{'s}$ and also includes all other overhead. Refer to Section III for further details. Best viewed in electronic format.}
\label{arch}
\end{center}
\end{figure*}

Commercial plenoptic cameras, like Lytro \cite{ng2005light}, have a lenslet array that helps capture both the intensity and direction of incident light rays. A popular approach to visualizing such 4D light field data is using the two-plane model, $L(u,v,s,t)$ \cite{levoy1996light}. In this, a 4D light field can be represented as a 2D array of 2D views or sub-aperture images (SAIs). The SAIs have a spatial dimension of $s\times t$ and exhibit parallax across the horizontal, $u$ and vertical, $v$ directions (Fig. \ref{arch} (a)). Despite the resulting spatial and angular variations within and across the SAIs, there is a high overlap in their content. This inspires the possibility of achieving a higher compression ratio by jointly compressing the SAIs rather than each individually. Consequently, light field image compression is motivated by the goal of taking advantage of this correlation within and among the SAIs.

In recent years, the learning-based 2D image compression field has been progressing by substantial strides \cite{balle2016end,theis2017lossy,balle2018variational}. A drawback of learning-based models, compared with ubiquitous standards like JPEG, could be its data dependency, that is, the inability of a single trained model to generalize across different test image formats. However, the capability to learn the specific transformation functions suitable to the ever-evolving complex imaging modalities is appealing \cite{hornik1989multilayer}. In addition, compared to hand-engineered codecs which take a considerable amount of time in getting standardized, learning-based approaches have been producing encouraging results in much shorter intervals of time. 
 
In this work, we extend the end-to-end learning-based approach to light field image compression. We propose a 3DCNN autoencoder with a disparity estimation component that ensures structural fidelity. The model is end-to-end trainable and does not depend on any separate hand-engineered modules. In contrast to most of the existing learning-based solutions, our model does not require any hand-crafted features for learning the disparity. The proposed model jointly optimizes for rate and distortion, thereby learning a better bit allocation scheme resulting in superior visual quality. Experimental results show that the model achieves superior performance compared to the state-of-the-art hand-engineered codecs in terms of the MS-SSIM metric. Though the model does not exhibit gains over the state-of-the-art in terms of PSNR, competitive performance shows its potential. Furthermore, compared to the existing hybrid codecs, the low-complexity design of the proposed model facilitates faster decoding using parallel processing.
  
 The rest of the paper is organized as follows. Section II provides a brief overview of prior work in the field, Section III presents the proposed method, Section IV provides the experimental results and analysis, and Section V concludes the paper.
 
 \section{Prior work}

Several light field image compression methods have been proposed over the years \cite{9020136}. Particularly for lossy compression of 4D light fields, pseudo-sequence-based methods using traditional video codecs is one of the commonly employed approaches \cite{liu2016pseudo}. Another popular approach involves dividing the SAIs into two sets, one of which is usually sparse. This is followed by the prediction of the other set of SAIs from the sparse set \cite{8030107,zhao2017light,viola2018graph}. The sparse set and prediction parameters are usually encoded using standard codecs, while various SAI prediction methods are used for assistance. In another direction, \cite{8022889} proposes encoding a low-rank approximation of the light field, while \cite{choudhury2015low} uses coded snapshots and sparse coding with a learned dictionary. Furthermore, the JPEG committee launched the JPEG Pleno initiative \cite{ebrahimi2016jpeg} for standardizing the compression of plenoptic data, which includes light fields, leading to a number of proposed solutions including \cite{astola2018wasp,de20184d}.  

With the advent of deep learning, there have been efforts in the direction of learning-based light field image compression as well \cite{hou2018light,bakir2018light,jia2018light,bakir2020light,liu2021view}. Commonly, the SAIs are divided into two sets, one of which is encoded using a standard video codec, and CNN or GAN-based models are used to predict the other set. The CNN-based angular super-resolution model of \cite{kalantari2016learning}, which uses a set of hand-crafted features for learning the disparity, forms a part of most of these approaches. While all the above learning-based approaches have a hybrid structure with dependence on multiple components, \cite{zhong20203d} proposes a simple 3DCNN autoencoder. The 3D convolution enables learning in the \textit{pseudo-temporal} dimension and hence the model has an input data structure awareness.
 
 \section{Proposed method}
The following section outlines the structuring of input data, followed by a brief operational and architectural overview, and finally, presents the proposed joint rate-distortion optimization.

\subsection{Input data structuring}

A number of SAIs can be extracted from the lenslet images such as ones obtained from a Lytro Illum camera. Most commonly they are decoded to $14\times14$ or $15\times15$ collection of SAIs. We decode the lenslet images to  $14\times14$, that is, a total of $196$ SAIs using the Light Field Toolbox \cite{dansereau2013decoding}. However, to avoid vignetting artifacts only the center $8\times8$ SAIs are used. This collection of SAIs can be represented by $\boldsymbol{X} \in  \mathbb{R}_{}^{8\times8\times h \times w \times 3}$, where $h$ and $w$ depend on the spatial resolution of individual (RGB) SAIs. Further, the SAIs are grouped row-wise. Only one such row is processed at a time, that is, the input is ${\boldsymbol{x}} \in \mathbb{R}_{}^{8\times h \times w \times 3}$ (Fig. \ref{arch} (a)). Thus, the 4D data is essentially converted into 3D.

\subsection{Operational overview}

The proposed model has multiple components to capture the light field data structure and also achieve compression while  maintaining an acceptable visual quality. However, the unified model can be thought of as an autoencoder that takes as its input, ${\boldsymbol{x}}$ as described above. The encoder $Enc$ performs a nonlinear parametric analysis transform to produce a latent representation ${{\boldsymbol{y}}} = Enc({\boldsymbol{x}},  \boldsymbol{\Theta})$. A quantization operation $Q[\cdot]$ is carried out in this latent space. Entropy coding of the quantized vectors $\boldsymbol{\hat{y}}=Q[{\boldsymbol{y}}]$ using a prior $p$ produces the bitstream $\boldsymbol{b}=\log_{2}(p(\boldsymbol{\hat{y}}))$. The encoder and decoder both have access to the prior. This enables the decoder $Dec$ to decode the quantized representation $\boldsymbol{\hat{y}}$ back from the bitstream. Finally, $Dec$ performs a nonlinear parametric synthesis transform to recover the reconstruction ${\boldsymbol{\hat{x}}} = Dec(\boldsymbol{\hat{y}},  \boldsymbol{\Phi})$. We train the model by optimizing the tradeoff between the length of the bitstream compared to the input (bitrate) and obtaining a faithful reconstruction of the input, that is, the rate-distortion tradeoff.

\subsection{Architectural overview}

From our experiments, a naive 3DCNN autoencoder for achieving ${\boldsymbol{\hat {x}}} = Dec(Q[Enc({\boldsymbol{x}} ,   \boldsymbol{\Theta})], \boldsymbol{\Phi})$, with $\boldsymbol{x}$ structured as above, seems to be incapable of learning the complex relationship among the SAIs. This is in part due to the lack of vertical disparity information in $\boldsymbol{x}$. Taking cues from the breakthrough work of \cite{kalantari2016learning}, our architecture has two main components: the first is a colour module and the second is composed of eight auxiliary disparity modules (Fig. \ref{arch} (c)). Each module has an encoder-decoder structure with a bottleneck to create latent representations of their inputs. The higher number of disparity modules facilitates learning of features as opposed to requiring feeding hand-crafted features in \cite{kalantari2016learning}. 

\subsection{Rate-distortion optimization}

The color module receives input ${\boldsymbol{x}} \in \mathbb{R}_{}^{8\times h \times w \times 3}$. Simultaneously, each $i_{}^{th}$ auxiliary disparity module receives feature tensor $\boldsymbol{f_{}^{i}}  \in \mathbb{R}_{}^{4\times h \times w \times 3}$ made up of $\boldsymbol{x}(i)$ (the  $i_{}^{th}$  SAI of the input row), $\boldsymbol{x}(c)$ (the center SAI of the input row), and $(u,v)$, the position indices of $\boldsymbol{x}(i)$ with respect to the central SAI of the $8\times8$ light field $\boldsymbol{X}$. The position indices $(u, v)$ help in learning the rigid disparity structure of input light field which is a function of the relative positions of the SAIs \cite{8030107}. The color module's encoder performs an analysis transform ${{\boldsymbol{y}}} = g_{a}({\boldsymbol{x}}, \boldsymbol{\theta_{g}})$, which is followed by a quantization operation. A prior, $p_{c}$, is used to entropy code the quantized latent vector $\boldsymbol{\hat{y}}=Q[{{\boldsymbol{y}}}]$. The prior is a fully factorized density model \cite{balle2018variational}. Similarly, each $i_{}^{th}$ auxiliary disparity module performs an analysis tranform to obtain latent representation $\boldsymbol{z_{}^{i}} = h_{a}^{i}(\boldsymbol{f_{}^{i}},  \boldsymbol{\theta_{h}^{i}})$. Another entropy model, $p_{d}$, similar to 
$p_{c}$ is used to entropy code the quantized disparity vectors $\boldsymbol{\hat{z}_{}^{i}}=Q[\boldsymbol{z_{}^{i}}]$. The decoder sides of the color module and each $i_{}^{th}$ auxiliary module then perform synthesis transforms to get the intermediate reconstruction, ${\boldsymbol{\tilde{x}}} = g_{s}(\boldsymbol{\hat{y}}, \boldsymbol{\phi_{g}}) $ and $i_{}^{th}$ disparity map, $\boldsymbol{d_{}^{i}} =  h_{s}^{i}(\boldsymbol{\hat{z}_{}^{i}}, \boldsymbol{\phi_{h}^{i}})$, respectively.  
 Each output disparity map $\boldsymbol{d_{}^{i}}$  is used to warp the corresponding 2D $\boldsymbol{\tilde{x}}(i)$ slice of the intermediate reconstructed tensor ${\boldsymbol{\tilde{x}}}$  to output the final reconstruction $\boldsymbol{\hat{x}}$. The warping operation uses bilinear interpolation for predicted disparities that are not integers. 
 
 Our aim here is to obtain reconstructions of the input subject to acceptable amounts of degradation, which is measured by a distortion loss $D$. In addition, we need to control the number of bits required for the encoded representation, which is given by the bitrate $R$.  This formulation gives us the overall objective to be optimized as $L=R+\lambda D$, where $\lambda$ is the Lagrangian multiplier. Using the cross-entropies with respect to $p_{l}$ (the marginal distribution of the latent vectors) as an estimation of the rate, $R$ and MSE as the distortion loss, D, we have:
 
\begin{multline}
L( \boldsymbol{\Theta,\Phi,\Psi}) =  \mathop{\mathbb{E}}_{{\boldsymbol{l}}\sim p_{{\boldsymbol{l}}}}[-\log_{2} p_{c}(Q[ g_{a}({\boldsymbol{x}},  \boldsymbol{\theta_{g}})]) + \\ \sum_{i=1}^{8}-\log_{2} p_{d}(Q[ h_{a}^{i}(\boldsymbol{f}_{}^{i},  \boldsymbol{\theta_{h}^{i}})])] +  \lambda \mathop{\mathbb{E}}_{{\boldsymbol{x}}\sim p_{{\boldsymbol{x}}}} \|{\boldsymbol{\hat{{\boldsymbol{x}}}-{\boldsymbol{x}}}}\|_{2}^{2} \label{eq} 
\end{multline}
 where, $p_{\boldsymbol{x}}$ is the unknown marginal distribution of input $\boldsymbol{x}$. The analysis parameters $( \boldsymbol{\theta_{g},\theta_{h}^{1}},...,\boldsymbol{\theta_{h}^{8}})$ are encapsulated in $ \boldsymbol{\Theta}$ and synthesis parameters  $( \boldsymbol{\phi_{g},\phi_{h}^{1}},...,\boldsymbol{\phi_{h}^{8}})$ in $ \boldsymbol{\Phi}$, while $ \boldsymbol{\Psi}$ represents the parameters of the priors $p_{c}$ and $p_{d}$. To facilitate training using gradient descent methods, the nondifferentiable quantization operation $Q[\cdot]$ is replaced by a mix of uniform noise \cite{balle2016end} and rounding \cite{theis2017lossy} as described in \cite{minnen2020channel}.

\section{Experiments and results}

In this section, we provide the implementation details and performance evaluations.  

\subsection{Datasets}

Our training dataset consists of $310$ light field images: $86$ from the EPFL Light field dataset\cite{rerabek2016new}, $72$ from \cite{kalantari2016learning}, $28$ from the Stanford Lytro dataset \cite{raj2016stanford}, $19$ from the synthetic 4D Light Field dataset \cite{honauer2016dataset} and $105$ from our own dataset created using a Lytro Illum camera. For training, we extract $64\times64$ patches from each SAI with a stride of $16$, while maintaining the row-wise input structure. This gives us around $940,000$ training samples. For testing, we use $18$ light field images: $9$ each from \cite{rerabek2016new} and \cite{kalantari2016learning}. For both training and testing, we use only the center $8\times8$ SAIs as described earlier. Also, there is no overlap between the train and test sets. 

\subsection{Implementation}

We implement our model using the Tensorflow framework. The color module's synthesis transform is composed of four layers of alternating strided 3D convolution layers and GDN nonlinearity which has been shown to perform well in finding compressive representations \cite{balle2016end}. The downsampling is achieved by using a stride of 2 at each convolutional step. The synthesis transform is designed as a corresponding reverse with downsampling replaced by upsampling and GDN by an approximate inverse, IGDN. The auxiliary disparity modules have a similar structure except three instead of four layers on each encoder and decoder sides. A kernel support of $3$ is used for the 3DCNN layers. The Adam optimizer \cite{kingma2014adam} is employed with default settings $\beta_1=0.9, \beta_2=0.999,\epsilon=1e_{}^{-7}$, a starting learning rate of $1e_{}^{-4}$ and a batch size of $30$. We train for different values of $\lambda$ to obtain different bitrates. The bottleneck channel dimension of the color module is one of $190, 320, 512 \text{ or } 720$, whereas the auxiliary modules have a bottleneck channel dimension of $8$. Warping with disparity maps is accomplished using bilinear interpolation.

\begin{table}[t]
\caption{Average Processing Time}
\begin{center}

\begin{tabular}{|c|c|c|c|c|c|c|c|}
\hline
\multicolumn{8}{|c|}{{Average encoding time (s)}} \\
\hline
{{{VVC}}}&  \multicolumn{2}{|c|}{{ \cite{bakir2020light}}} & { \cite{jia2018light}} &  { \cite{hou2018light}} &  { \cite{zhong20203d}}& \multicolumn{2}{|c|}{ {Proposed}}  \\
\hline
{CPU}&{CPU} &{GPU} &{CPU} &{CPU} & & {CPU}&{GPU} \\
\hline
$145$&$445$&$335$&$291$&$6028$& $100$&$\mathbf{22}$&$\mathbf{14}$\\
\hline
\multicolumn{8}{|c|}{{Average decoding time (s)}} \\
\hline
$\boldsymbol{4}$&$124$&$94$&$53$&$583$&&$14$&$\mathbf{12}$\\
\hline
\end{tabular}

\label{tab1}
\end{center}
\end{table}

\subsection{Evaluation}

\subsubsection{Metric}

The performance is evaluated in terms of the average luminance PSNR distortion metric as recommended in \cite{convenorjpeg}, and the average luminance MS-SSIM (in terms of dB), denoted as PSNR\_Y and MS-SSIM\_Y, respectively. The bitrate is calculated using bits per pixel (bpp). 

It is worth noting that for the proposed model, the bpp includes all overheads such as the shape of the tensors (required for decoding), as well as the encoded disparity maps. 

\subsubsection{Methods for comparison} 

We compare our model to the state-of-the-art hand-engineered codecs,  \begin {enumerate*} [label=\itshape\alph*\upshape)] \item the OpenJPEG \cite{oj} implementation of JPEG2000 is used with default settings, \item the HEVC reference software HM-16.22 \cite{hm} is used with profile Main and low delay configuration, and \item the VVC reference software VTM-11.0 \cite{vtm}  is used with profile Auto and low delay configuration. \end {enumerate*} 

For the video codecs, the RGB SAIs are converted into planar YUV 420 videos and encoded in serpentine order. Data is obtained for four different Quantization Parameters (QPs). Correspondingly, data is obtained from the proposed model trained for four different $\lambda$ values. 
\begin{figure}[t]
\centering
\subcaptionbox{}{\includegraphics[width=0.4\textwidth]{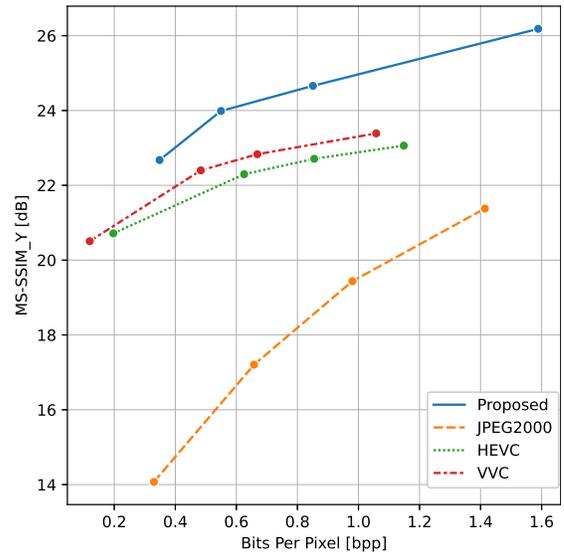}}\hfill
\subcaptionbox{}{\includegraphics[width=0.4\textwidth]{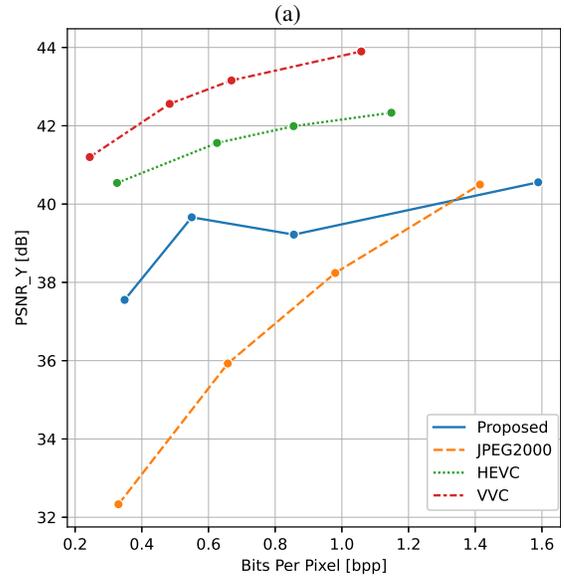}}%
\caption{Rate-distortion curves (a)  MS-SSIM\_Y-bitrate (b) PSNR\_Y-bitrate comparisons with standard codecs averaged over 18 test light field scenes.}
\label{figcomp}
\end{figure}

\subsubsection{Results} 
Fig. \ref{figcomp} (a) shows MS-SSIM\_Y as a function of bitrate. The proposed model outperforms all three standard codecs. In terms of PSNR\_Y (Fig. \ref{figcomp} (b)), the proposed method achieves a BD-BR saving of $-16.26$ \% and BD-PSNR gain of $+2.86$ dB over JPEG2000 \cite{bjontegaard2001calculation}, but underperforms compared to the video codecs. The inconsistent performance of the proposed model at the higher bitrate could be due to insufficient training for the particular $\lambda$ values. Again, in terms of RGB PSNR, the proposed model outperforms both VVC and HEVC. Table \ref{tab2} shows RGB PSNR values averaged across the 4 lenslet images available in the JPEG Pleno Light Field dataset \cite{jpegpdb}. Comparing with the only other end-to-end learning-based method \cite{zhong20203d} in terms of the reported RGB PSNR values there seems to be no advantage. However, there is a gain in terms of the reported encoding speed. 

Table \ref{tab1} shows a comparison of processing times of various existing solutions. For our method, the value represents the time required when using parallel processing for each row of input light field. The performance is tested on a system with Intel Core i7-6700 3.40GHz, 32GB memory, and NVIDIA Quadro RTX 6000 GPU. Compared to the existing learning-based solutions, the proposed method outperforms in terms of both the encoding and decoding speeds. Though VVC has a superior decoding speed owing to its efficient hand-engineered design, the proposed model has a higher encoding speed which may be desirable for some applications. 

Training a similar model without the auxiliary disparity module results in blurrier reconstructions with disparity being captured in only the $u$ dimension and not in the $v$ dimension. Similarly, a model with the same architecture as proposed, except 2DCNN layers in place of 3DCNN in the color module, produces reconstructions with artifacts like missing pixels at occlusions. This happens due to the encoder not being able to capture the color information of the entire input light field.

It is important to note that we trained our model for much lower than 1M steps as compared to 5M steps in \cite{minnen2020channel} with similar models for 2D images. At the same time, our training dataset is much smaller as compared to \cite{zhong20203d} which uses more than 6000 light field images and a patch size of $256\times256$ for training. The loss used is simple MSE as compared to a perceptual loss in \cite{liu2021view}. This shows the model has many possible directions for improvement. 
  \begin{table}[t]
\caption{Avg. RGB PSNR-Bitrate}
\begin{center}

\begin{tabular}{|c|c|c|c|c|}
\hline
&BPP1&PSNR1&BPP2&PSNR2 \\
\hline
HEVC&$0.67$ &$32.8$&$0.92$& $33.07$ \\
\hline
VVC&$0.72$&$33.49$&${0.99}$&$33.73$\\
\hline
Proposed&$\boldsymbol{0.63}$&$\boldsymbol{33.62}$&$\boldsymbol{0.86}$&$\boldsymbol{33.84}$\\
\hline
\end{tabular}

\label{tab2}
\end{center}
\end{table}

\section{Discussion}

The proposed model learns the structural information of light fields without requiring ground truth disparity maps or hand-crafted features. The model is generalisable to different light field image formats, limited only by the availability of training data. Since the model optimises for not only pixel-to-pixel distortion but also the structure, it may encourage a higher MSE and lower PSNR values. This is, however, only a hypothesis and we have not performed extensive evaluations in this direction. Apart from objective tests, subjective evaluations need to be carried out to judge if the structure is well captured. Overall, we show that this is a viable research direction that requires further exploration and targeted efforts to develop optimal architectures. In future, we plan on extending the model to accept the entire 4D light field to increase the compression performance. Further, the idea of \textit{hyperpriors} \cite{balle2018variational}, which is an instantiation of side-information, can be incorporated.
\bibliographystyle{IEEEtran}
\bibliography{IEEEabrv,bibfile}

\begin{thebibliography}{10}
\providecommand{\url}[1]{#1}
\csname url@samestyle\endcsname
\providecommand{\newblock}{\relax}
\providecommand{\bibinfo}[2]{#2}
\providecommand{\BIBentrySTDinterwordspacing}{\spaceskip=0pt\relax}
\providecommand{\BIBentryALTinterwordstretchfactor}{4}
\providecommand{\BIBentryALTinterwordspacing}{\spaceskip=\fontdimen2\font plus
\BIBentryALTinterwordstretchfactor\fontdimen3\font minus
  \fontdimen4\font\relax}
\providecommand{\BIBforeignlanguage}[2]{{%
\expandafter\ifx\csname l@#1\endcsname\relax
\typeout{** WARNING: IEEEtran.bst: No hyphenation pattern has been}%
\typeout{** loaded for the language `#1'. Using the pattern for}%
\typeout{** the default language instead.}%
\else
\language=\csname l@#1\endcsname
\fi
#2}}
\providecommand{\BIBdecl}{\relax}
\BIBdecl

\bibitem{ng2005light}
R.~Ng, M.~Levoy, M.~Br{\'e}dif, G.~Duval, M.~Horowitz, and P.~Hanrahan, ``Light
  field photography with a hand-held plenoptic camera,'' Ph.D. dissertation,
  Stanford University, 2005.

\bibitem{levoy1996light}
M.~Levoy and P.~Hanrahan, ``Light field rendering,'' in \emph{Proceedings of
  the 23rd Annual Conference on Computer Graphics and Interactive Techniques},
  1996, pp. 31--42.

\bibitem{balle2016end}
J.~Ball{\'e}, V.~Laparra, and E.~P. Simoncelli, ``End-to-end optimized image
  compression,'' in \emph{5th International Conference on Learning
  Representations, (ICLR)}, 2017.

\bibitem{theis2017lossy}
L.~Theis, W.~Shi, A.~Cunningham, and F.~Husz{\'{a}}r, ``Lossy image compression
  with compressive autoencoders,'' in \emph{5th International Conference on
  Learning Representations, (ICLR)}, 2017.

\bibitem{balle2018variational}
J.~Ball{\'e}, D.~Minnen, S.~Singh, S.~J. Hwang, and N.~Johnston, ``Variational
  image compression with a scale hyperprior,'' in \emph{International
  Conference on Learning Representations, (ICLR)}, 2018.

\bibitem{hornik1989multilayer}
K.~Hornik, M.~Stinchcombe, and H.~White, ``Multilayer feedforward networks are
  universal approximators,'' \emph{Neural networks}, vol.~2, no.~5, pp.
  359--366, 1989.

\bibitem{9020136}
C.~{Conti}, L.~D. {Soares}, and P.~{Nunes}, ``Dense light field coding: A
  survey,'' \emph{IEEE Access}, vol.~8, pp. 49\,244--49\,284, 2020.

\bibitem{liu2016pseudo}
D.~Liu, L.~Wang, L.~Li, Z.~Xiong, F.~Wu, and W.~Zeng, ``Pseudo-sequence-based
  light field image compression,'' in \emph{IEEE International Conference on
  Multimedia \& Expo Workshops (ICMEW)}, 2016, pp. 1--4.

\bibitem{8030107}
J.~{Chen}, J.~{Hou}, and L.~{Chau}, ``Light field compression with
  disparity-guided sparse coding based on structural key views,'' \emph{IEEE
  Transactions on Image Processing}, vol.~27, no.~1, pp. 314--324, 2018.

\bibitem{zhao2017light}
S.~Zhao and Z.~Chen, ``Light field image coding via linear approximation
  prior,'' in \emph{IEEE International Conference on Image Processing
  (ICIP)}.\hskip 1em plus 0.5em minus 0.4em\relax IEEE, 2017, pp. 4562--4566.

\bibitem{viola2018graph}
I.~Viola, H.~P. Maretic, P.~Frossard, and T.~Ebrahimi, ``A graph learning
  approach for light field image compression,'' in \emph{Applications of
  Digital Image Processing XLI}, vol. 10752.\hskip 1em plus 0.5em minus
  0.4em\relax International Society for Optics and Photonics, 2018, p. 107520E.

\bibitem{8022889}
X.~{Jiang}, M.~{Le Pendu}, R.~A. {Farrugia}, and C.~{Guillemot}, ``Light field
  compression with homography-based low-rank approximation,'' \emph{IEEE
  Journal of Selected Topics in Signal Processing}, vol.~11, no.~7, pp.
  1132--1145, 2017.

\bibitem{choudhury2015low}
C.~Choudhury, Y.~Tarun, A.~Rajwade, and S.~Chaudhuri, ``Low bit-rate
  compression of video and light-field data using coded snapshots and learned
  dictionaries,'' in \emph{IEEE 17th International Workshop on Multimedia
  Signal Processing (MMSP)}, 2015, pp. 1--6.

\bibitem{ebrahimi2016jpeg}
T.~Ebrahimi, S.~Foessel, F.~Pereira, and P.~Schelkens, ``{JPEG Pleno: Toward an
  efficient representation of visual reality},'' \emph{IEEE Multimedia},
  vol.~23, no.~4, pp. 14--20, 2016.

\bibitem{astola2018wasp}
P.~Astola and I.~Tabus, ``{WaSP: Hierarchical warping, merging, and sparse
  prediction for light field image compression},'' in \emph{2018 7th European
  Workshop on Visual Information Processing (EUVIP)}.\hskip 1em plus 0.5em
  minus 0.4em\relax IEEE, 2018, pp. 1--6.

\bibitem{de20184d}
M.~B. de~Carvalho \emph{et~al.}, ``{A 4D DCT-based lenslet light field
  codec},'' in \emph{2018 25th IEEE International Conference on Image
  Processing (ICIP)}.\hskip 1em plus 0.5em minus 0.4em\relax IEEE, 2018, pp.
  435--439.

\bibitem{hou2018light}
J.~Hou, J.~Chen, and L.-P. Chau, ``Light field image compression based on
  bi-level view compensation with rate-distortion optimization,'' \emph{IEEE
  Transactions on Circuits and Systems for Video Technology}, vol.~29, no.~2,
  pp. 517--530, 2018.

\bibitem{bakir2018light}
N.~Bakir, W.~Hamidouche, O.~D{\'e}forges, K.~Samrouth, and M.~Khalil, ``Light
  field image compression based on convolutional neural networks and linear
  approximation,'' in \emph{25th IEEE International Conference on Image
  Processing (ICIP)}, 2018, pp. 1128--1132.

\bibitem{jia2018light}
C.~Jia, X.~Zhang, S.~Wang, S.~Wang, and S.~Ma, ``Light field image compression
  using generative adversarial network-based view synthesis,'' \emph{IEEE
  Journal on Emerging and Selected Topics in Circuits and Systems}, vol.~9,
  no.~1, pp. 177--189, 2018.

\bibitem{bakir2020light}
N.~Bakir, W.~Hamidouche, S.~A. Fezza, K.~Samrouth, and O.~D{\'e}forges, ``Light
  field image coding using dual discriminator generative adversarial network
  and {VVC} temporal scalability,'' in \emph{IEEE International Conference on
  Multimedia and Expo (ICME)}, 2020, pp. 1--6.

\bibitem{liu2021view}
D.~Liu, X.~Huang, W.~Zhan, L.~Ai, X.~Zheng, and S.~Cheng, ``View
  synthesis-based light field image compression using a generative adversarial
  network,'' \emph{Information Sciences}, vol. 545, pp. 118--131, 2021.

\bibitem{kalantari2016learning}
N.~K. Kalantari, T.-C. Wang, and R.~Ramamoorthi, ``Learning-based view
  synthesis for light field cameras,'' \emph{ACM Transactions on Graphics
  (TOG)}, vol.~35, no.~6, pp. 1--10, 2016.

\bibitem{zhong20203d}
T.~Zhong, X.~Jin, and K.~Tong, ``{3D-CNN} autoencoder for plenoptic image
  compression,'' in \emph{IEEE International Conference on Visual
  Communications and Image Processing (VCIP)}, 2020, pp. 209--212.

\bibitem{dansereau2013decoding}
D.~G. Dansereau, O.~Pizarro, and S.~B. Williams, ``Decoding, calibration and
  rectification for lenselet-based plenoptic cameras,'' in \emph{Proceedings of
  the IEEE Conference on Computer Vision and Pattern Recognition, (CVPR)},
  2013, pp. 1027--1034.

\bibitem{minnen2020channel}
D.~Minnen and S.~Singh, ``Channel-wise autoregressive entropy models for
  learned image compression,'' in \emph{2020 IEEE International Conference on
  Image Processing (ICIP)}, 2020, pp. 3339--3343.

\bibitem{rerabek2016new}
M.~Rerabek and T.~Ebrahimi, ``New light field image dataset,'' in \emph{8th
  International Conference on Quality of Multimedia Experience (QoMEX)}, 2016.

\bibitem{raj2016stanford}
\BIBentryALTinterwordspacing
A.~S. Raj, M.~Lowney, R.~Shah, and G.~Wetzstein, ``Stanford lytro light field
  archive,'' 2016. [Online]. Available:
  \url{http://lightfields.stanford.edu/LF2016.html}
\BIBentrySTDinterwordspacing

\bibitem{honauer2016dataset}
K.~Honauer, O.~Johannsen, D.~Kondermann, and B.~Goldluecke, ``A dataset and
  evaluation methodology for depth estimation on {4D} light fields,'' in
  \emph{Asian Conference on Computer Vision}.\hskip 1em plus 0.5em minus
  0.4em\relax Springer, 2016, pp. 19--34.

\bibitem{kingma2014adam}
D.~P. Kingma and J.~Ba, ``Adam: A method for stochastic optimization,''
  \emph{arXiv preprint arXiv:1412.6980}, 2014.

\bibitem{convenorjpeg}
{JPEG Convenor}, ``{JPEG} {P}leno call for proposals on light field coding,''
  \emph{Doc. ISO/IEC JTC1/SC29/WG1/N/74014}.

\bibitem{oj}
\BIBentryALTinterwordspacing
``{OpenJPEG}.'' [Online]. Available:
  \url{https://github.com/uclouvain/openjpeg}
\BIBentrySTDinterwordspacing

\bibitem{hm}
\BIBentryALTinterwordspacing
``{HM} reference software for {HEVC}.'' [Online]. Available:
  \url{https://vcgit.hhi.fraunhofer.de/jvet/HM}
\BIBentrySTDinterwordspacing

\bibitem{vtm}
\BIBentryALTinterwordspacing
``{VTM} reference software for {VVC}.'' [Online]. Available:
  \url{https://vcgit.hhi.fraunhofer.de/jvet/VVCSoftware_VTM}
\BIBentrySTDinterwordspacing

\bibitem{bjontegaard2001calculation}
S.~Matyunin, ``{Bjontegaard metric calculation (BD-PSNR) - ver 1.0},''
  \emph{MATLAB Central File Exchange}, 2013.

\bibitem{jpegpdb}
\BIBentryALTinterwordspacing
``{JPEG Pleno Light Field Datasets according to common test conditions}.''
  [Online]. Available: \url{http://plenodb.jpeg.org/lf/pleno_lf}
\BIBentrySTDinterwordspacing

\end{thebibliography}

\end{document}